\documentclass[journal,onecolumn,11pt]{IEEEtran}
\usepackage{amssymb, cite,dblfloatfix, amsmath,amsthm}
\usepackage{graphicx}
\usepackage{color}
\usepackage{setspace}
\usepackage{floatrow}
\newtheorem{theorem}{\bf Theorem}

\title{Linear Predictability in MRI Reconstruction: Leveraging Shift-Invariant Fourier Structure for Faster and Better Imaging}
\author{Justin P. Haldar and Kawin Setsompop}

\begin{document}

\maketitle 

\bstctlcite{IEEE:BSTcontrol}

\doublespacing

Magnetic resonance imaging (MRI) is a powerful and highly-versatile imaging technique that has had a tremendous impact in both science and medicine.  Unfortunately, MRI data acquisition is also time consuming and expensive, which has thus far prevented it from delivering on its full potential.  As a result, the MRI field has always been interested in signal processing methods that can generate high-quality images from a small amount of measured data. These kinds of methods can increase the comfort of the person being scanned, enable higher-quality assessment of time-varying phenomena, improve scanner throughput, and/or allow more detailed and comprehensive MRI examinations within a fixed total imaging time.

Over the past several decades, many different computational approaches have been proposed for reducing scan time.  While it may not be widely known to either the broader MRI community or the broader signal processing community, linear prediction provides a powerful mathematical framework for understanding a wide range of existing computational MRI reconstruction methods.  In this paper, we provide an overview of such methods in the context of this framework.    Linear prediction is well known in signal processing \cite{vaidyanathan2008} and may be most recognizable for its usefulness in speech processing and spectrum estimation applications.  In MRI, linear predictability  implies that  data does not need to be sampled as often as dictated by the conventional sampling theorem, since the missing data may be accurately imputed as a linear combination of measured samples.

Linear prediction underlies some of the earliest methods in the computational MRI reconstruction field \cite{smith1986,liang1989,martin1989,liang1992,sodickson1997}, some of the most widely utilized computational MRI reconstruction methods in modern clinical practice \cite{griswold2002,setsompop2012}, and some of the most flexible and versatile modern computational imaging approaches that are enabling unprecedented new styles of data acquisition \cite{shin2014,haldar2013b}.  In addition, the concept of linear predictability can be used to unify a number of more classical MRI reconstruction constraints \cite{haldar2013b,shin2014,haldar2015,jin2015a,ongie2016,haldar2015b}, including limited spatial support and smooth phase constraints \cite{liang1992}, multi-channel (``parallel imaging'') constraints \cite{ying2010}, and sparsity constraints \cite{lustig2008}.  Importantly, this can be done without needing to make the strong discrete/finite-dimensional image modeling assumptions of typical model-based MRI reconstruction methods \cite{fessler2010b} and without requiring the detailed prior information or calibration data that is frequently required by classical constrained reconstruction methods \cite{liang1992,ying2010}.  

Significantly, while many MRI reconstruction methods are implicitly tied to linear prediction, this connection has not always been explicit, and a unified view of these methods under a common theoretical framework has emerged only recently.  The connection to linear prediction is theoretically illuminating and has a number of practical benefits.  For example, it can help demystify the performance characteristics of early approaches that were originally presented as ``black-boxes'' without strong theoretical justification, helping to reveal both the strengths and limitations of such methods.  The new insights that  emerge from this perspective  can also be used to identify areas where the inherent structure of MRI data is currently underexploited and inspire creative new approaches that can help bring MRI closer to its as-yet-undetermined fundamental performance limits.  This is an active area of research, and we believe there are substantial opportunities for further development.

\section{The Basic MRI Model}\label{sec:basicmri}

MRI images are generally multidimensional.  However, to avoid complicated notation, our description will focus on a simplified version of MRI in which we want to reconstruct a continuous one-dimensional complex-valued image function $\rho(x)$ with $x\in\mathbb{R}$.  Generalizations to higher-dimensional scenarios are straightforward.   The  image is assumed to have finite spatial support, such that $\rho(x) = 0$ for $x \notin [-B/2,B/2]$.  The variable $B$ defines an upper bound on the size of the image support, and the interval $[-B/2,B/2]\subset \mathbb{R}$ is known as the \emph{Field of View} (FOV).  

Classical MRI acquisition is usually modeled as sampling the Fourier transform of $\rho(x)$ at the Nyquist rate, i.e.,
\begin{equation}
\tilde{\rho}[n] = \int_{-B/2}^{B/2} \rho(x) e^{-i2\pi n x/B} dx,\label{eq:int}
\end{equation}
where $\tilde{\rho}[n]$ is the $n$th sample in the Fourier domain, and conventional sampling theory tells us that we can recover the original image from infinite samples via
\begin{equation}
\rho(x) = \frac{1}{B} \sum_{n=-\infty}^\infty \tilde{\rho}[n] e^{i 2\pi n x/B}
\end{equation}
for $x \in [-B/2,B/2]$.  Unfortunately, MRI data acquisition time scales with the number of samples of $\tilde{\rho}[n]$ that are measured, while image quality is severely degraded if too few samples are acquired.

\section{Linear Predictive Extrapolation,  Interpolation, and Annihilation}

Fortunately, for many MRI images of interest, the Fourier data $\tilde{\rho}[n]$ is approximately linearly predictable because it is well-modeled as autoregressive.\footnote{Note that some early work assumed an autoregressive moving average (ARMA) model for $\tilde{\rho}[n]$ \cite{smith1986}, although this has fallen out of favor relative to pure autoregressive modeling, and this article will focus purely on the autoregressive case.}  Linear predictability can be expressed in many forms, although the form that is potentially most familiar to a signal processing audience is what we will call the \emph{extrapolation form} in this work.

\subsection{Linear Predictive Extrapolation}
The extrapolation form assumes that there exists a fixed shift-invariant set of $P$ coefficients $\alpha_1,\alpha_2,\ldots,\alpha_P$ such that, for all integers $n$, the sample $\tilde{\rho}[n]$ can be approximated as a linear combination of the past $P$ samples, i.e.,
\begin{equation}
\tilde{\rho}[n] \approx \sum_{k=1}^P \alpha_k \tilde{\rho}[n-k] \text{ for } \forall n\in\mathbb{Z}.\label{eq:extrap}
\end{equation}
This relationship is valuable, because if the $\alpha_k$ coefficients can be estimated, then it can be possible to recursively extrapolate an arbitrary number of samples of $\tilde{\rho}[n]$ from as few as $P$ consecutive measurements!  This type of extrapolation procedure has been used in the early constrained MRI literature to achieve super-resolution, i.e., generating a high-resolution image by extrapolating unmeasured high-resolution information content from low-resolution measured data \cite{liang1989,smith1986,martin1989,liang1992}, and was used for similar purposes even earlier in magnetic resonance spectroscopy (MRS) under subtly different theoretical principles \cite{barkhuijsen1985,koehl1999}.\footnote{The theory of linear prediction for MRS assumes that the measured data can be represented as a linear combination of exponentially-decaying complex sinusoids.  While decaying sinusoids are perfectly linearly predictable, their Fourier transforms are not bandlimited. Bandlimitedness is central to the theoretical arguments we present herein, and this article does not attempt to cover linear prediction theory for signals that are not-bandlimited.  Readers interested in this topic are referred to a review article on linear prediction for MRS \cite{koehl1999}.}  While extrapolation approaches can work reasonably well in certain circumstances, extrapolation over long-distances can be very sensitive to noise and modeling errors \cite{liang1992}, which has limited the widespread practical deployment of these methods in MRI reconstruction.

\subsection{Linear Predictive Interpolation}

\begin{figure}[tp]
\floatbox[{\capbeside\thisfloatsetup{capbesideposition={right,center},capbesidewidth=2.5in}}]{figure}[\FBwidth]{\caption{Illustrative example of the interpolation form of linear prediction for the 2D case.  In this example, the central sample (red) is approximated as a linear combination of the neighboring samples (blue) from a 2D circular  neighborhood \cite{haldar2013b}.\label{fig:interp}}}{\includegraphics[width=1.5in]{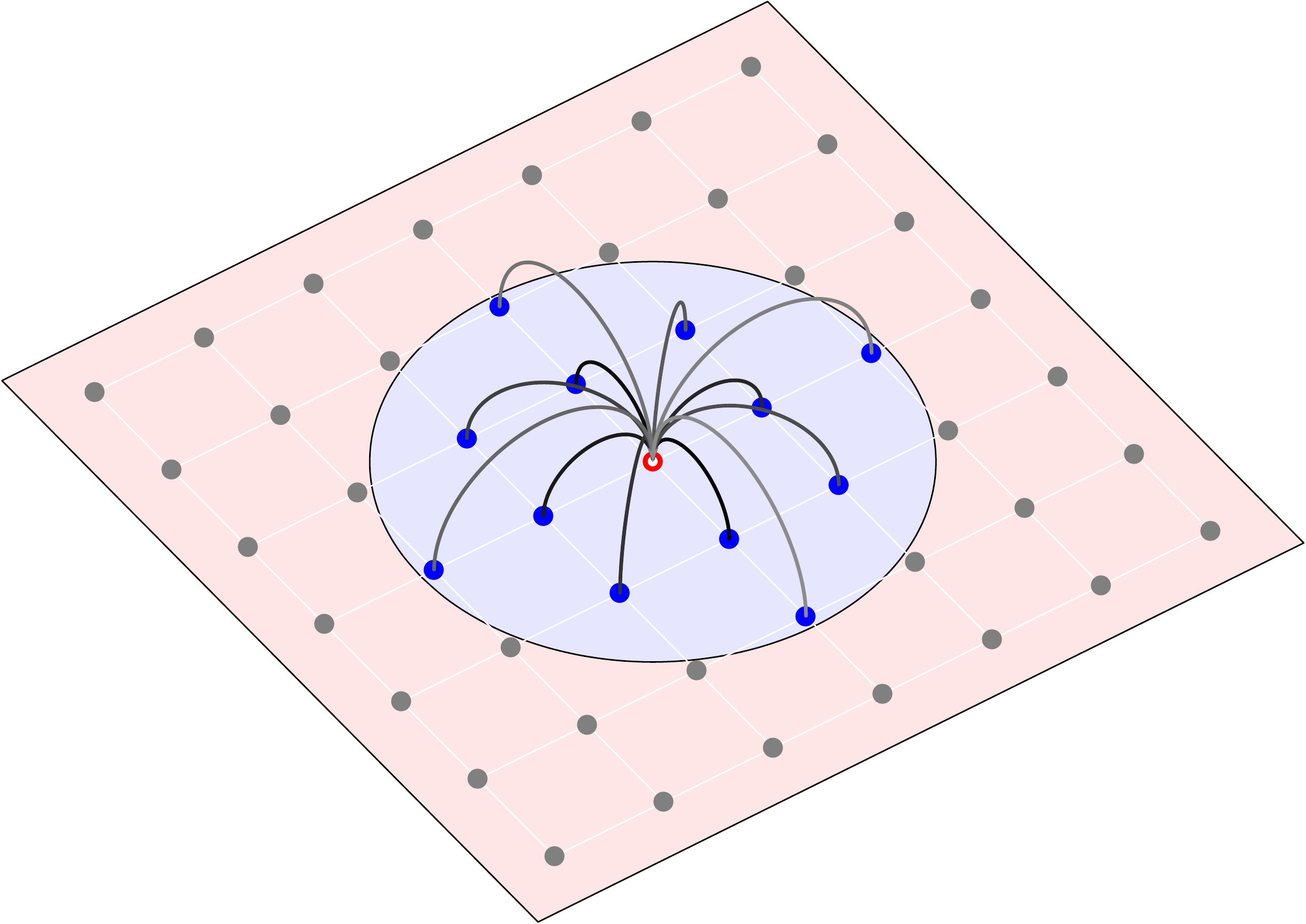}}
\end{figure}

A more recent development is to use linear prediction to perform interpolation \cite{dologlou1996,griswold2002,lustig2010}, which is less sensitive to noise and modeling errors than extrapolation.  The \emph{interpolation form} assumes that, for all integers $n$, the sample $\tilde{\rho}[n]$ can be approximated as a linear combination of both past and future samples, e.g.,
\begin{equation}
\tilde{\rho}[n] \approx \sum_{k=-L}^{-1} \alpha_{k} \tilde{\rho}[n-k] + \sum_{k=1}^{P} \alpha_{k} \tilde{\rho}[n-k] \text{ for } \forall n\in\mathbb{Z}.\label{eq:interp}
\end{equation}
This type of linear predictability would allow MRI images to be accurately and stably reconstructed from high-resolution data that is sampled below the Nyquist rate.  Although Eq.~\eqref{eq:interp} is written for the one-dimensional case, a graphical illustration of higher-dimensional linear predictive interpolation is presented in Fig.~\ref{fig:interp}.

\subsection{Linear Prediction and Annihilation}

If the signal $\tilde{\rho}[n]$ is approximately linearly predictable in the sense of  Eq.~\eqref{eq:extrap} or Eq.~\eqref{eq:interp}, then it is easy  to see that there must exist coefficients $\{\tilde{h}[n]\}_{n=-L}^{P}$ with $\tilde{h}[0]=-1$, such that
\begin{equation}
0 \approx \sum_{k=-L}^{P} \tilde{h}[k] \tilde{\rho}[n-k] \text{ for } \forall n\in\mathbb{Z}.\label{eq:annihil}
\end{equation}
In particular, Eq.~\eqref{eq:extrap} is obtained by taking $L=0$, $\tilde{h}[0]=-1$, and $\tilde{h}[n]=\alpha_n$ for $n=1,\ldots,P$, while Eq.~\eqref{eq:interp} is obtained by taking $\tilde{h}[0]=-1$ and $\tilde{h}[n]= \alpha_n$ for $n\in\left\{-L,\ldots,P\right\}\setminus \{0\}$.

The relationship in Eq.~\eqref{eq:annihil} implies that $P+L+1$ consecutive samples are approximately linearly dependent, such that any one missing sample can be predicted as the weighted sum of the others \cite{cheung1990}.  This relationship is also shift-invariant and takes the form of a convolution. As a result, Eq.~\eqref{eq:annihil} can be called an approximate \emph{annihilating filter relationship} \cite{blu2008} because the signal $\tilde{\rho}[n]$ is being approximately annihilated by convolution with the ``filter'' function $\tilde{h}[n]$.  Linear predictability is thus equivalent to the existence of a non-trivial approximate annihilating filter.

\section{Computational Image Reconstruction with Linear Prediction}\label{sec:recon}

\begin{figure}[htp]
\begin{minipage}{7in}
\centering
\begin{minipage}{0.8in}
\centering 
\footnotesize {\color{white}.} Full {\color{white}.} Sampling \\[0.25em]
\includegraphics[width=.8in]{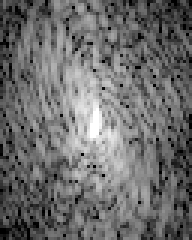}
\end{minipage}
\begin{minipage}{0.8in}
\centering 
\footnotesize Low {\color{white}p}Resolution{\color{white}p}  \\[0.25em]
\includegraphics[width=.8in]{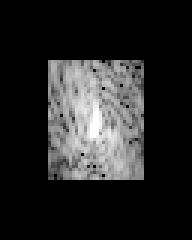}
\end{minipage}
\begin{minipage}{0.8in}
\centering 
\footnotesize Uniform 1D (w/calibration) \\[0.25em]
\includegraphics[width=.8in]{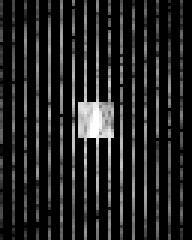}
\end{minipage}
\begin{minipage}{0.8in}
\centering 
\footnotesize Random (w/calibration) \\[0.25em]
\includegraphics[width=.8in]{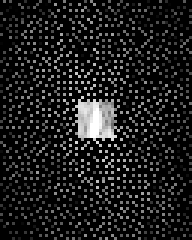}
\end{minipage}
\begin{minipage}{0.8in}
\centering 
\footnotesize Random PF (w/calibration) \\[0.25em]
\includegraphics[width=.8in]{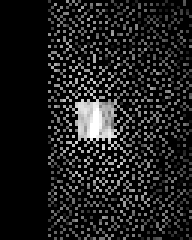}
\end{minipage}
\begin{minipage}{0.8in}
\centering 
\footnotesize {\color{white}..} Random {\color{white}..} (no calibration) \\[0.25em]
\includegraphics[width=.8in]{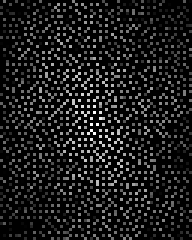}
\end{minipage}
\begin{minipage}{0.8in}
\centering 
\footnotesize Random PF (no calibration) \\[0.25em]
\includegraphics[width=.8in]{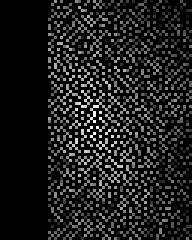}
\end{minipage}
\begin{minipage}{0.8in}
\centering 
\footnotesize Unconventional (USC logo) \\[0.25em]
\includegraphics[width=.8in]{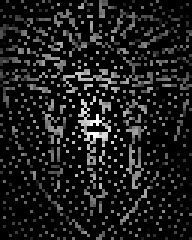}
\end{minipage}
\end{minipage}
\caption{This figure shows illustrative sampling patterns for 2D Fourier data.  The leftmost image shows the fully sampled data, while the remaining images show zero-filled undersampled datasets that can all be reconstructed with various linear prediction-based approaches.}\label{fig:sampling}
\end{figure}

Before attempting to justify the assumption of linear predictability, we will first start by describing some of the prevailing approaches for exploiting this assumption when it is applicable.  The use of linear predictability in computational MRI reconstruction has evolved over several decades, and with each new technical innovation, the power and flexibility of these techniques has continued to grow.  This has allowed the range of Fourier sampling patterns that are compatible with this kind of reconstruction to  increase over time.  Since different sampling strategies will play a role in the story, we illustrate a few 2D examples in Fig.~\ref{fig:sampling} that we will refer to in what follows. 

\subsection{Extrapolation}
In the early days \cite{smith1986,liang1989,liang1992,martin1989}, linear prediction ideas were exclusively used to achieve high-resolution reconstruction from low-resolution data, a version of the common \emph{super-resolution} problem.  This was often  achieved by collecting a large consecutive set of samples of $\tilde{\rho}[n]$ (often called \emph{calibration data} in modern MRI terminology), and then using that calibration data to estimate a single set of coefficients $\{\alpha_k\}_{k=1}^P$ that yields good prediction performance within the calibration region.   Interestingly, this type of approach can be viewed as an early example of a data-driven ``learning" approach where the calibration region serves as training data. Once estimated, the linear prediction coefficients can then be used outside of the calibration region to extrapolate missing high-resolution information based on repeated iterations of Eq.~\eqref{eq:extrap}.  Since MRI images have most of their energy concentrated at low frequencies, calibration regions are usually chosen within the low-resolution part of the Fourier domain, as illustrated in Fig.~\ref{fig:sampling}.  

\subsection{Interpolation}

After extrapolation, one of the next big innovations was the idea to perform interpolation of undersampled data (i.e., data sampled below the Nyquist rate) \cite{dologlou1996,griswold2002,lustig2010}. The most prominent example is GRAPPA  \cite{griswold2002}, which is widely deployed by commercial MRI scanner vendors.\footnote{Methods like GRAPPA \cite{griswold2002} and SPIRiT \cite{lustig2010} were originally introduced in the context of reconstructing multi-channel images acquired simultaneously in parallel using an array of receiver coils.  However, our description in this section adopts single-channel notation for the sake of simplicity, although we will expound upon the multi-channel setting in Sec.~\ref{sec:single}.} GRAPPA  performs linearly predictive interpolation of undersampled k-space using a relationship like Eq.~\eqref{eq:interp}, where a single-set of interpolative prediction coefficients is estimated from a fully-sampled set of calibration data.   However, one of the challenges in using this kind of interpolation relationship is that, in order to interpolate a single missing value of $\tilde{\rho}[n]$, Eq.~\eqref{eq:interp} seems to require the availability of $P+L$ neighboring samples $\tilde{\rho}[n-k]$ for appropriate values of $k$.  Interpolation would offer very limited benefits if we were required to sample $P+L$ samples for every sample we omit, since this would not allow high undersampling factors.  However, methods like GRAPPA were able to overcome this issue by estimating linear interpolation relationships that are aware of the local sampling pattern, and which, e.g., estimate the prediction coefficients for $\tilde{\rho}[n]$ while forcing $\alpha_k$ to be zero for each value of $k$ for which $\tilde{\rho}[n-k]$ was not acquired \cite{griswold2002}.  

While popular, this approach also has a few limitations.  On one hand, it is necessary to estimate a different set of linear prediction coefficients for each distinct local sampling pattern, and this could be computationally burdensome if there are a large number of distinct local sampling patterns.  As a result, uniform sampling (cf. Fig.~\ref{fig:sampling}) is the most popular sampling approach for methods like GRAPPA. Another challenge is due to the fact that, while it may be easy to find annihilation relationships (as will be discussed in Sec.~\ref{sec:single}), it can be substantially more difficult to find good linear prediction relationships if the undersampling factor is high and many of the $\alpha_k$ coefficients are  restricted to be zero to account for unsampled neighbors.  

\subsection{Annihilation}

One of the next innovations is exemplified by SPIRiT \cite{lustig2010}, which replaces the use of direct linear predictive interpolation with the use of a single annihilation relationship in the form of Eq.~\eqref{eq:annihil}.  SPIRiT uses calibration data to estimate an annihilation relationship, and then solves an inverse problem that penalizes inconsistency with the annihilation relationship.  Specifically, given annihilation coefficients $\{\tilde{h}[n]\}$, the SPIRiT approach is equivalent to finding coefficients $\{\tilde{\rho}[n]\}$ that minimize
\begin{equation}
\sum_{n } \left|\sum_{k=-L}^{P} \tilde{h}[k] \tilde{\rho}[n-k]\right|^2
\end{equation}
subject to data consistency constraints, where, without loss of generality, we have used the notation of the single-image case.  As mentioned before, one advantage of this approach over interpolation-based reconstruction is that annihilation relationships can be shown to exist under much weaker conditions, and there is no need to artificially force the values of certain  prediction coefficients to zero. Another advantage is that, since this approach effectively imposes linear predictability constraints using regularization, it is straightforward to also involve other regularization penalties in the reconstruction to also enforce other constraints (e.g.,  the $\ell_1$-norm to enforce sparsity \cite{lustig2008}).  The main downside of this annihilation approach relative to  interpolation is increased computational complexity.

While the original SPIRiT just uses a single annihilation relationship, later methods like PRUNO \cite{zhang2011} use a similar approach but leveraging a multiplicity of different annihilating filters instead of just one.  The rationale for the existence of multiple annihilating filters will be described later in Sec.~\ref{sec:single}.

\subsection{Phase Constraints and Partial Fourier Methods}

Another important innovation is the realization that smooth image phase constraints could be used to allow one side of the Fourier domain to be  linearly predicted from data on the opposite side.  To gain some intuition for this approach, consider a real-valued image with zero-phase.  In this toy example, the symmetry property of the Fourier transform tells us that Fourier samples will possess conjugate symmetry, and therefore that missing data from one side of the Fourier domain can be trivially reconstructed from the complementary sample on the opposite side.  MRI images almost always have nontrivial spatial phase variations in real applications, but in these circumstances, a missing sample can still often be estimated from samples on the opposite side if the image phase can be estimated accurately \cite{liang1992}.  This leads to a data acquisition strategy that is frequently termed \emph{Partial Fourier} (PF) acquisition, in which one side of the Fourier domain is not included in the Fourier sampling pattern, while low-frequency calibration data is acquired to enable estimation of the image phase.  While phase-constrained PF reconstruction has existed in the MRI literature for a long time \cite{liang1992}, the incorporation of this idea into the linear prediction framework is a more recent development that enables enhanced flexibility  \cite{huang2009,blaimer2009, haldar2013b, haldar2015, haldar2015b}, such as the ability to use phase constraints to improve reconstruction quality with both the PF and non-PF sampling schemes from Fig.~\ref{fig:sampling}.

\subsection{Calibrationless Reconstruction with Structured Low-Rank Matrix Modeling}

For the most part, the previously-described approaches all rely on calibration data to enable the pre-training of linear prediction coefficients or annihilating filters.  However, one of the most recent developments is the use of structured low-rank matrix completion methods to enable high-quality reconstruction even if calibration data has not been acquired (cf. Fig.~\ref{fig:sampling}).  In particular, it has been known for a long time  that if MRI data is linearly predictable, then a Hankel or Toeplitz matrix formed from that data is expected to have approximately low-rank \cite{liang1989, zhang2011} and that enforcing low-rank structure on such a matrix may enable the imputation of missing data \cite{dologlou1996}.  From a modern view \cite{haldar2013b}, the convolutional structure of a Hankel or Toeplitz matrix means that  a distinct nullspace vector will exist for each annihilating filter that applies to the data. The existence of multiple distinct annihilating filters will thus imply that the matrix must approximately have low rank. This enables the use of reconstruction methods that leverage low-rank matrix completion \cite{shin2014, haldar2013b, haldar2015, haldar2015b, ongie2016, jin2015a,jacob2019}. One of the nice features of this approach is that, while annihilating filters can exist for a number of reasons (as will be described in Sec.~\ref{sec:single}), the low-rank modeling approach will implicitly identify every applicable annihilation relationship, without requiring prior knowledge that the image obeys a specific constraint.  Instead, the nullspace of the matrix will implicitly capture all of the relevant constraints, without the need for any user intervention.  This makes it relatively safe to, e.g.,  choose a formulation that is capable of simultaneously imposing annihilation relationships associated with multiple constraints (e.g., the limited support constraints, smooth phase constraints, and parallel imaging constraints that will be described in Sec.~\ref{sec:single}), even if the user is not sure whether all of these constraints will apply.  However,  high computational complexity is one of the main drawbacks of structured low-rank matrix completion relative to the other approaches.

While we will not cover the details of structured low-rank matrix recovery (which is the topic of another article in this special issue \cite{jacob2019}), we will still make some relevant historical                                                                                                                               comments.  Early structured low-rank matrix approaches  \cite{liang1989,dologlou1996} considered simple one-dimensional image reconstruction problems, and this low dimensionality was a major limitation for the practical performance of such approaches.  However, multi-dimensional versions of structured low-rank matrix recovery have recently been enabled by modern improvements in computational power to impose parallel imaging constraints \cite{shin2014, haldar2015, jin2015a}, support and phase constraints \cite{haldar2013b}, and transform domain sparsity constraints \cite{ongie2016, jin2015a, haldar2015b}.  Among other things, these methods are enabling a range of unprecedented calibrationless sampling strategies \cite{shin2014, haldar2013b,haldar2015,haldar2015b}.   This increased flexibility is perhaps best exemplified by the fact that it is now possible to reconstruct high-quality images from data sampled in very unconventional ways, like sampling patterns designed based on the logo of the University of Southern California (USC) \cite{haldar2015b} which do not sample the low-frequency Fourier information very densely (cf. Fig.~\ref{fig:sampling}).  While we do not suggest that this type of unconventional sampling is optimal in any way, we also do not know what type of sampling strategy is actually optimal for this setting, because these new approaches have disrupted the conventional wisdom in this area.  There remain a large number of unanswered questions related to optimal sampling design, with ample opportunities for future research.  Nevertheless, since linear predictive relationships rely on local Fourier information, a good rule of thumb for sampling design is that sparsely sampled Fourier regions are usually more difficult to reconstruct than densely sampled ones, except perhaps in multi-image cases if a region that was sampled sparsely for one dataset has been sampled more densely in one of the other  datasets.

\subsection{An Example Reconstruction Illustration}

\begin{figure}[htp]
\centering
\includegraphics[width=\textwidth]{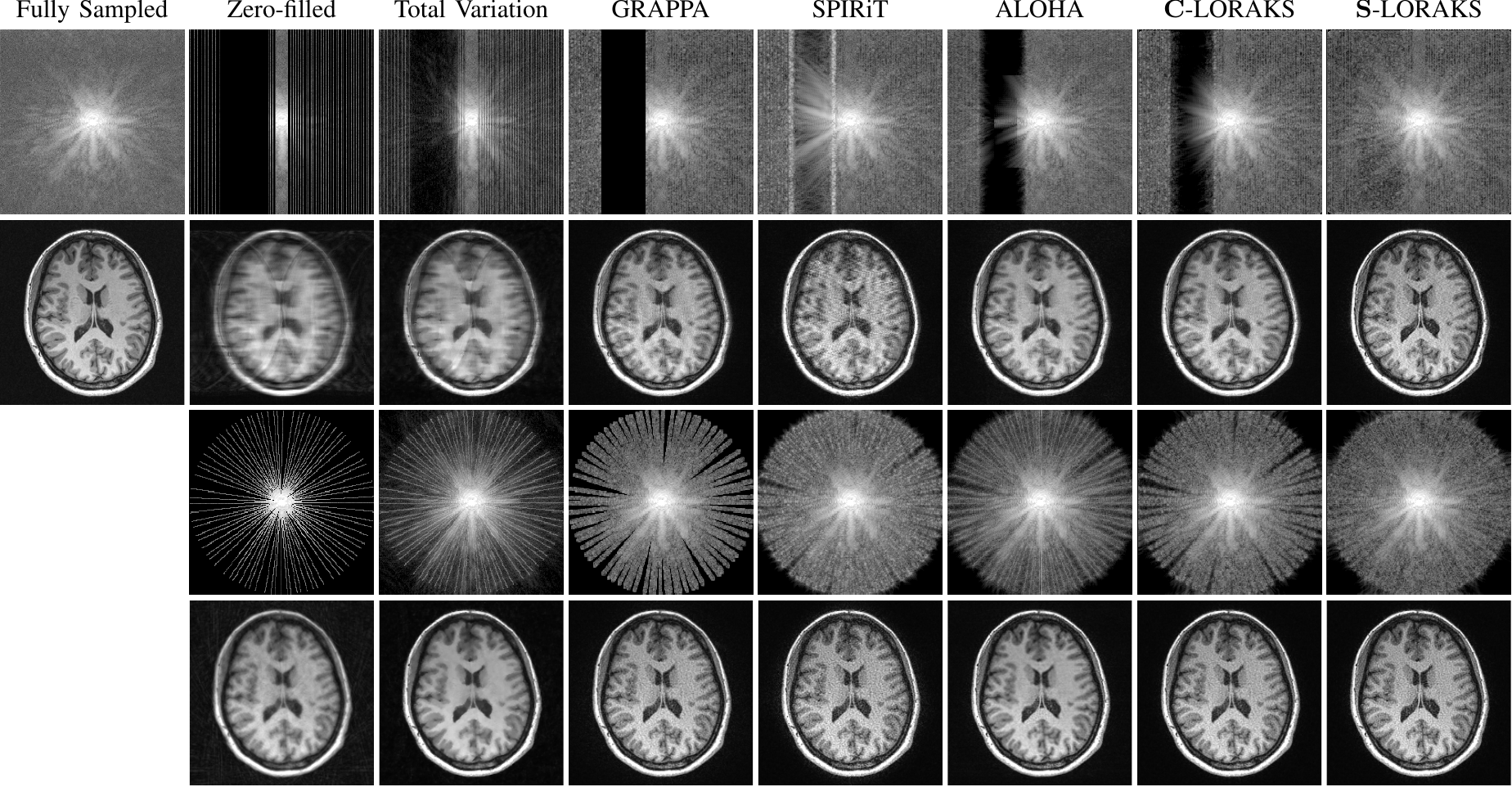}
\caption{Illustration of two reconstruction scenarios in which 25\% of the data (an acceleration factor of 4) is sampled using a 12-channel array of receiver coils and different sampling patterns.  Top two rows: uniform 1D undersampling with calibration data and a large gap.  Bottom two rows: a golden angle pseudo-radial sampling pattern. While calibration data is available from the fully-sampled center of k-space in both cases, the associated reconstruction problems are still challenging  because of substantial gaps in the k-space coverage, as can be seen in zero-filled data. For each sampling pattern, the top row shows 2D Fourier coefficients, while the bottom row shows  corresponding images.  GRAPPA \cite{griswold2002} reconstruction uses interpolation relationships estimated from the calibration data to impute missing data, while SPIRiT \cite{lustig2010} uses a single annihilation relationship estimated from the calibration data.  The ALOHA \cite{jin2015a} and LORAKS \cite{haldar2015} methods use multiple annihilation relationships within a structured low-rank matrix recovery formulation.  $\mathbf{C}$-based LORAKS assumes support and parallel imaging constraints \cite{haldar2013b,haldar2015,haldar2015b}, $\mathbf{S}$-based LORAKS assumes support, phase, and parallel imaging constraints \cite{haldar2013b,haldar2015,haldar2015b}, and ALOHA assumes parallel imaging and transform-domain sparsity constraints \cite{jin2015a}.  As can be seen in these examples, there is a clear difference in reconstruction characteristics as we move from interpolation (GRAPPA) to annihilation (SPIRiT) to multiple annihilation relationships (ALOHA and LORAKS).  For reference, we also show a sparsity-inducing regularization strategy as often used for compressed sensing  \cite{lustig2008} (in this case, a joint total variation (TV) roughness penalty across all channels). It is clearly apparent that this type of sparsity constraint used by itself is less powerful than many of the linear prediction-based approaches in this example, although the performance of sparsity constraints can be easily improved by incorporating additional information about image phase and parallel imaging \cite{block2007}. In addition, while our illustration showed the use of different constraints in isolation to highlight their distinct characteristics, it should also be mentioned that it is often straightforward to combine multiple constraints together to achieve even higher performance (e.g., \cite{lustig2008,shin2014, haldar2013b, haldar2015}). }   \label{fig:recon}
\end{figure}

To wrap up this section we show some quick examples in Fig.~\ref{fig:recon} that illustrate the changes in reconstruction performance that have accompanied evolutions in linear prediction-based reconstruction strategies.   Note that most quantities (i.e., the image $\rho(x)$, data $\tilde{\rho}[n]$, etc.) will be complex-valued in practical MRI applications, although this figure (and most subsequent figures, unless otherwise specified) only depicts magnitudes for simplicity. 

This illustration is not intended to provide a head-to-head comparison between state-of-the-art methods, since state-of-the art methods will frequently combine multiple approaches together.  Rather, this example is intended to convey some of the relative characteristics of different approaches when they are used in isolation. It should also be noted that relative performance of different methods can vary substantially depending on the image characteristics and the specific sampling pattern that was used, which means the relative ranking of different methods implied by Fig.~\ref{fig:recon} may also vary from one scenario to the next. Importantly,  all of the linear prediction-based reconstruction methods in this figure are available through open-source software  \cite{uecker2015,aloha2017,kim2018b}, which allows readers to experiment for themselves with these different approaches in different contexts.

\section{When is $\tilde{\rho}[n]$ linearly predictable?}\label{sec:single}

The previous section showed that linear predictability can be a powerful practical tool for MRI image reconstruction.  In this section, we describe some of the theoretical underpinnings for linear predictability.

\subsection{Linear Predictability for a Single Image} \label{sec:sing}
To start, we will consider the basic imaging setup from Sec.~\ref{sec:basicmri}, and address the fundamental question: under what situations will Eq.~\eqref{eq:annihil} hold? It turns out that this type of linear predictability will occur if and only if the effective spatial support  of $\rho(x)$ possesses certain characteristics.  In particular, consider the following theorem, which represents a formal statement of  arguments from Refs.~\cite{cheung1990,haldar2013b}:
\begin{theorem}
Assume that $\tilde{\rho}[n]$ is defined as in Eq.~\eqref{eq:int}, and let $\varepsilon$ be an arbitrary positive  scalar. We have that $$\sum_{n \in \mathbb{Z}} \left|\sum_{k=-L}^{P} \tilde{h}[k] \tilde{\rho}[n-k]\right|^2 \leq \varepsilon $$ for some  filter function $\tilde{h}[n]$ if and only if  $$ \frac{1}{B}\int_{-B/2}^{B/2} |\rho(x) h(x)|^2 dx \leq \varepsilon, \mathrm{\, where \,}
h(x) = \frac{1}{B} \sum_{n=-L}^{P} \tilde{h}[n] e^{i 2\pi n x/B}.$$
\end{theorem}
This theorem is a natural consequence of Parseval's theorem combined with the convolution theorem of the Fourier transform.  

Theorem~1 provides both necessary and sufficient conditions for  approximate linear predictability  in the form of Eq.~\eqref{eq:annihil} (with approximation quality measured in the $\ell_2$-norm of infinite sequences).  In particular, Eq.~\eqref{eq:annihil} requires that a function $h(x)$ exists  such that $\rho(x) h(x) \approx 0$ within the FOV  \cite{haldar2013b} (with approximation quality measured in the $\mathcal{L}_2$ norm of continuous functions).  Moreover, $h(x)$ must be smooth because it is represented as a bandlimited Fourier series, and must also have large energy (i.e., we must have that $\int_{-B/2}^{B/2} |h(x)|^2 dx \geq B$ since $\tilde{h}[0] = -1$).

Achieving $\rho(x)h(x)\approx0$ requires that, for almost every spatial location $x$ within the FOV, either $\rho(x)\approx 0$ or $h(x)\approx 0$.  Since our smoothness and largeness conditions on $h(x)$ prevent the possibility of $h(x) \approx 0$ almost everywhere, we should only expect to observe good linear predictability if $\rho(x) \approx 0$ in at least some subregion of the FOV.  In other words, the existence of good k-space linear prediction relationships is intimately connected to whether  the support (or approximate support) of the image is smaller than the FOV \cite{haldar2013b, cheung1990}.  Most MRI images do have support that is strictly smaller than the FOV -- for example, most body parts are more ellipsoidal than rectangular,  so that the corners of an MRI image are often empty for rectangular FOVs.  This suggests that most MRI images will possess some degree of linear predictability!  

\subsubsection{Choosing good $h(x)$ functions}

If $\rho(x)$ is known, then the theory for constructing good $h(x)$ functions is intimately connected to both Slepian's classic work on discrete prolate spheroidal sequences (which are finite discrete sequences that have maximal energy  concentration on a specified subregion of the Fourier domain) \cite{slepian1978} and Cheung's  work on annihilation-based interpolation of bandlimited functions  \cite{cheung1990}.  In particular, define $\mathcal{G}$ as the self-adjoint linear operator that  maps input sequences  $\{\tilde{a}[n]\}_{n=-L}^P$ to output sequences $\{\tilde{b}[n]\}_{n=-L}^P$ according to
\begin{equation}
\tilde{b}[k] = \sum_{n=-L}^P \tilde{a}[n] \tilde{g}[n-k] \text{ for } k=-L,\ldots,P, \mathrm{\,   where  \,}
\tilde{g}[n] = \frac{1}{B^2} \int_{-B/2}^{B/2} |\rho(x)|^2 e^{i2\pi x n /B} dx.
\end{equation}
Following an argument similar to Slepian \cite{slepian1978}, it is easy to show  that if $\{\tilde{h}[n]\}_{n=-L}^P$ is a unit-normalized eigensequence  associated with the smallest eigenvalue of $\mathcal{G}$, then the function $h(x) =  \mathcal{F}(\{\tilde{h}[n]\}_{n=-L}^P)$ has its spatial-domain energy optimally distributed away from the energy of $\rho(x)$.  In particular, this choice of $\{\tilde{h}[n]\}_{n=-L}^P$ yields an $h(x)$ that will minimize $ \int_{-B/2}^{B/2} |\rho(x) h(x)|^2 dx$ within the class of unit-norm functions with appropriately-bandlimited Fourier series. 

While this approach to designing $h(x)$ may be attractive, it is also important to note that the construction of $\mathcal{G}$ requires perfect prior knowledge of the infinite dimensional image $|\rho(x)|$, which will not be available in practical applications.  As a result, most practical reconstruction methods will use alternative approaches to estimate annihilation functions from a finite amount of measured data (e.g., calibration data), as discussed previously.  Nevertheless, there is significance to the observation that a good annihilation filter is associated with a small-eigenvalue eigensequence (i.e., an approximate nullspace vector) of a certain linear operator -- in particular, the structured low-rank matrix modeling methods we described previously rely on very similar nullspace concepts!

\subsubsection{Multiplicity of $h(x)$ functions}

While many of the earliest linear-predictive MRI reconstruction methods \cite{smith1986,martin1989,liang1992,sodickson1997,mckenzie2001,griswold2002,setsompop2012,lustig2010} only made use of a single annihilation relationship  (i.e., a single filter $\{\tilde{h}[n]\}_{n=-L}^P$), it has more recently emerged \cite{zhang2011, haldar2013b, haldar2015,ongie2016, jin2015a, shin2014,haldar2015b} that there are frequently multiple distinct filters that all do a good job of approximately annihilating the data. This is beneficial because the inverse problem associated with computational image reconstruction from highly-undersampled data is often less ill-posed when multiple linear prediction relationships are utilized instead of just one.  In particular, ill-posedness occurs because at high undersampling factors, the space of images $\rho(x)$ that are consistent with the measured data can be large, while each linear prediction relationship can individually be interpreted as constraining the image $\rho(x)$ to approximately lie within a specific subspace associated with the filter coefficients.  Using multiple linear prediction relationships simultaneously can be interpreted as constraining the image to approximately lie within the intersection of these individual subspaces.  This has the effect of significantly reducing the amount of potential ambiguity in the solution to the inverse problem.

\begin{figure}[htp]
\floatbox[{\capbeside\thisfloatsetup{capbesideposition={right,center},capbesidewidth=3.5in}}]{figure}[\FBwidth]{
\caption{An illustration of a typical MRI image $\rho(x)$ (left) together with a collection of multiple distinct appropriately-bandlimited $h(x)$ functions (in red, with an  edge map for the original image provided  for spatial reference in white) that all have spatial-domain energy that is concentrated outside the support of $\rho(x)$.  As a result, these functions all satisfy $\rho(x)h(x) \approx 0$, and therefore all correspond to different approximate Fourier-domain linear prediction relationships. Corresponding annihilating filters $\tilde{h}[n]$ are also depicted below each $h(x)$.  \label{fig:annihil}}}{
\begin{minipage}{\linewidth}
\centering
\begin{minipage}{0.19\linewidth}
\centering 
\footnotesize $\rho(x)$ \\[0.25em]
\includegraphics[width=0.6in]{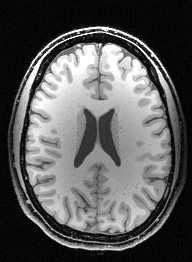}\\[0.25em]
\includegraphics[width=0.6in]{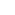}
\end{minipage}
\hspace{-0.15in}
\begin{minipage}{0.19\linewidth}
\centering 
\footnotesize $h_1(x)$\\[0.25em]
\includegraphics[width=0.6in]{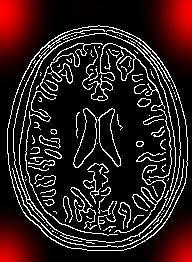}\\[0.25em]
\includegraphics[width=0.6in]{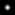}
\end{minipage}
\hspace{-0.15in}
\begin{minipage}{0.19\linewidth}
\centering 
\footnotesize $h_2(x)$ \\[0.25em]
\includegraphics[width=0.6in]{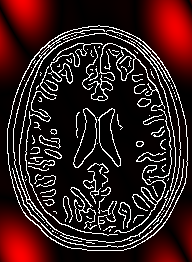}\\[0.25em]
\includegraphics[width=0.6in]{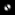}
\end{minipage}
\hspace{-0.15in}
\begin{minipage}{0.19\linewidth}
\centering 
\footnotesize $h_3(x)$\\[0.25em]
\includegraphics[width=0.6in]{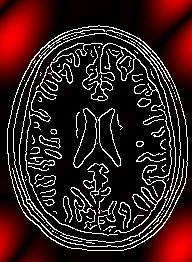}\\[0.25em]
\includegraphics[width=0.6in]{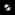}
\end{minipage}
\hspace{-0.15in}
\begin{minipage}{0.19\linewidth}
\centering 
\footnotesize $h_4(x)$\\[0.25em]
\includegraphics[width=0.6in]{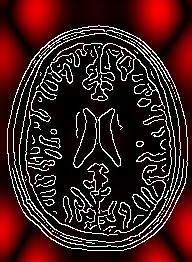}\\[0.25em]
\includegraphics[width=0.6in]{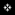}
\end{minipage}
\end{minipage}}
\end{figure}

From a theoretical perspective, a large multiplicity of approximate linear prediction relationships is able to occur in the present context because if $\rho(x)$ is support-limited, then there are often many linearly independent bandlimited functions $h(x)$ that satisfy $\rho(x)h(x) \approx 0$ \cite{cheung1990,haldar2013b}.  An orthonormal set of such functions $h(x)$ can be obtained by considering all of the eigensequences associated with the small eigenvalues of $\mathcal{G}$, which by construction are each associated with small values of $\int_{-B/2}^{B/2} |\rho(x) h(x)|^2 dx$.  An illustration of this is shown in Fig.~\ref{fig:annihil}.  

While most MRI images possess some form of approximate linear predictability, perfect linear predictability (with $\varepsilon=0$ in Thm. 1) is only available for a certain class of simple images that obey extremely stringent continuous-domain sparsity constraints.  In particular, perfect linear predictability requires that $\int_{-B/2}^{B/2}|\rho(x)h(x)|^2 dx$  is identically zero, but this is not generally possible unless the image $\rho(x)$ is supported on a set of measure zero.  (Note that as a nontrivial analytic function, $h(x)$ can only satisfy $h(x) =0$ on at most a set of spatial locations of measure zero -- as a result, $\rho(x)$ must be zero almost everywhere else!).   Previous work has developed deep theory that enables  perfect reconstruction of such images \cite{liang1989,ongie2016,jin2015a}, including, e.g., piecewise polynomial images (whose derivatives obey strict continuous-domain sparsity constraints \cite{liang1989}).  Our article will not focus on this case, as this theory is covered in another article in this special issue \cite{jacob2019}.  Note that for practical applications, there is generally not a significant difference between $\varepsilon =0$ and $\varepsilon \approx 0$ in Thm.~1, since real images  will rarely satisfy the highly-restrictive modeling assumptions required for $\varepsilon=0$ and because real data  will always contain noise.

\subsection{Linear Predictability and High-Pass Filtering}

\begin{figure}[htp]
\floatbox[{\capbeside\thisfloatsetup{capbesideposition={right,center},capbesidewidth=3.5in}}]{figure}[\FBwidth]{
\caption{An illustration of a high-pass filtered version $\mathcal{D}\rho(x)$ of the MRI image $\rho(x)$ from Fig.~\ref{fig:annihil}, which is approximately zero for a much more substantial region of the FOV than $\rho(x)$ was.  We also show a collection of multiple distinct appropriately-bandlimited $h(x)$ functions (in red, with an  edge map for the original image provided  for spatial reference in white) that have spatial-domain energy that is concentrated outside the support of $\mathcal{D}\rho(x)$, and therefore correspond to  different approximate Fourier-domain linear prediction relationships for appropriately weighted Fourier data $\tilde{w}[n]$. Corresponding annihilating  filters $\tilde{h}[n]$ are also depicted below each $h(x)$.  It should be noted that the annihilating filters from Fig.~\ref{fig:annihil} are generally also annihilating filters for this setting, since the support of $\mathcal{D}\rho(x)$ should be a subset of the support of $\rho(x)$. However, additional annihilation relationships are now available because $\mathcal{D}\rho(x)$ has  gaps in its support that were not available for $\rho(x)$.  Because these gaps are often narrow, the $h(x)$ functions in this case generally need to have higher bandwidth than those in Fig.~\ref{fig:annihil} to take advantage of the additional support characteristics of $\mathcal{D}\rho(x)$ beyond those inherited from $\rho(x)$.   \label{fig:annihil2}}}{
\begin{minipage}{\linewidth}
\centering
\begin{minipage}{0.19\linewidth}
\centering
\footnotesize $\mathcal{D}\rho(x)$ \\[0.25em]
\includegraphics[width=0.6in]{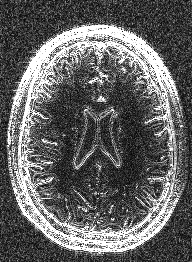}\\[0.25em]
\includegraphics[width=0.6in]{./graphics/fig2/foriginal.png}
\end{minipage}
\hspace{-0.15in}
\begin{minipage}{0.19\linewidth}
\centering 
\footnotesize $h_1(x)$\\[0.25em]
\includegraphics[width=0.6in]{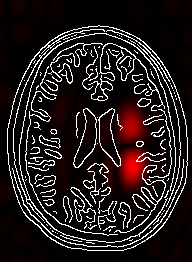}\\[0.25em]
\includegraphics[width=0.6in]{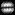}
\end{minipage}
\hspace{-0.15in}
\begin{minipage}{0.19\linewidth}
\centering 
\footnotesize $h_2(x)$ \\[0.25em]
\includegraphics[width=0.6in]{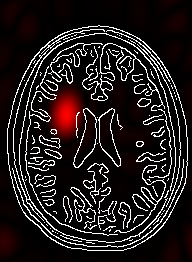}\\[0.25em]
\includegraphics[width=0.6in]{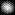}
\end{minipage}
\hspace{-0.15in}
\begin{minipage}{0.19\linewidth}
\centering 
\footnotesize $h_3(x)$\\[0.25em]
\includegraphics[width=0.6in]{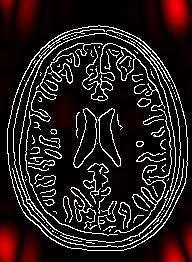}\\[0.25em]
\includegraphics[width=0.6in]{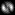}
\end{minipage}
\hspace{-0.15in}
\begin{minipage}{0.19\linewidth}
\centering 
\footnotesize $h_4(x)$\\[0.25em]
\includegraphics[width=0.6in]{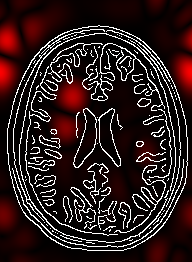}\\[0.25em]
\includegraphics[width=0.6in]{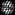}
\end{minipage}
\end{minipage}}
\end{figure}

Although the above theory covers the linear predictability of $\tilde{\rho}[n]$, it is interesting to note that many of the earliest linear-prediction based computational MRI reconstruction methods instead considered linear prediction of the Fourier data that would have been obtained from a high-pass filtered version of the image \cite{liang1989,liang1992,martin1989}.  While this type of data is not actually measured in MRI, it is easy to synthesize from $\tilde{\rho}[n]$ by using, e.g., the derivative property of the Fourier transform.  In particular, if $\tilde{\rho}[n]$ is the Fourier data corresponding to the image $\rho(x)$, then the sequence $\tilde{w}[n] = (i2\pi n/B) \tilde{\rho}[n]$ is the Fourier data corresponding to $\frac{\partial}{\partial x} \rho(x)$, where the spatial derivative operator $\frac{\partial}{\partial x}$ acts as a high-pass filter \cite{liang1989}.  The potential benefit of this data transformation is that $\frac{\partial}{\partial x} \rho(x)$ often has its energy concentrated near image edges.  This can make it easier to construct good annihilation functions $h(x)$, since  the effective support of $\frac{\partial}{\partial x} \rho(x)$ is generally much smaller than the effective support of $ \rho(x)$.  However, since the gaps over which $\frac{\partial}{\partial x} \rho(x)\approx 0$ tend to be relatively narrow because edge features are often found in close proximity to one another, it is often necessary to use $h(x)$ functions with much higher bandwidth (i.e., substantially larger values of $L$ and $P$) for this case compared to the previous case.    This is illustrated in Fig.~\ref{fig:annihil2}.

Notably, by applying linear prediction relationships to high-pass  data, these early reconstruction methods \cite{liang1989,liang1992,martin1989}  were effectively (and sometimes explicitly) relying on a continuous-domain concept of transform-domain sparsity constraints, and this was all done many years before the introduction and popularization of the modern ``compressed sensing'' approach to sparsity-constrained MRI \cite{lustig2008}! While the early methods applied these concepts to impose transform-domain sparsity in 1D settings for extrapolation \cite{liang1989,liang1992,martin1989}, some of the recent literature has applied these concepts in higher dimensions while considering more general sampling patterns \cite{ongie2016,jin2015a,haldar2015b}. 

Importantly, while most modern compressed sensing MRI approaches (e.g., those based on $\ell_1$-norm minimization \cite{lustig2008}) impose no particular structure on the sparsity pattern, this is not the same for linear predictive approaches which are implicitly associated with additional structural constraints.  In particular, the fact that the annihilating $h(x)$ functions are bandlimited requires that the support of the image (or its transform-domain representation) must have substantial gaps that are largely devoid of signal energy \cite{haldar2013b,haldar2015b}.  While such additional constraints on the support might not be universally applicable to every imaging scenario, they can yield improved reconstruction performance in the many practical applications where they do apply. This may help to explain the empirical performance advantages that have  been recently observed for linear-predictive methods over conventional compressed sensing approaches \cite{haldar2013b,jin2015a,ongie2016,jacob2019,haldar2015b,haldar2015}.

\subsection{Multiple images of the same anatomy}\label{sec:multi}
The theory in the previous sections gave necessary and sufficient conditions for the existence of approximate linear prediction relationships in the Nyquist-sampled Fourier data of a single image.  While these linear prediction relationships are already potentially useful, further  linear prediction relationships  can frequently be derived in scenarios where multiple correlated images are acquired of the same anatomy.  

The basic premise of this case is that datasets $\tilde{\rho}_q[n]_{q=1}^Q$ are acquired for a sequence of $Q$ images $\{\rho_q(x)\}_{q=1}^Q$.  In the context of the present article, we will assume that each of these images $\rho_q(x)$ represents a different modulation of the same underlying image $\rho(x)$, i.e., $\rho_q(x) = c_q(x) \rho(x)$ for $q=1,\ldots,Q$. 

\begin{figure}[htp]
\includegraphics[width=4.8in]{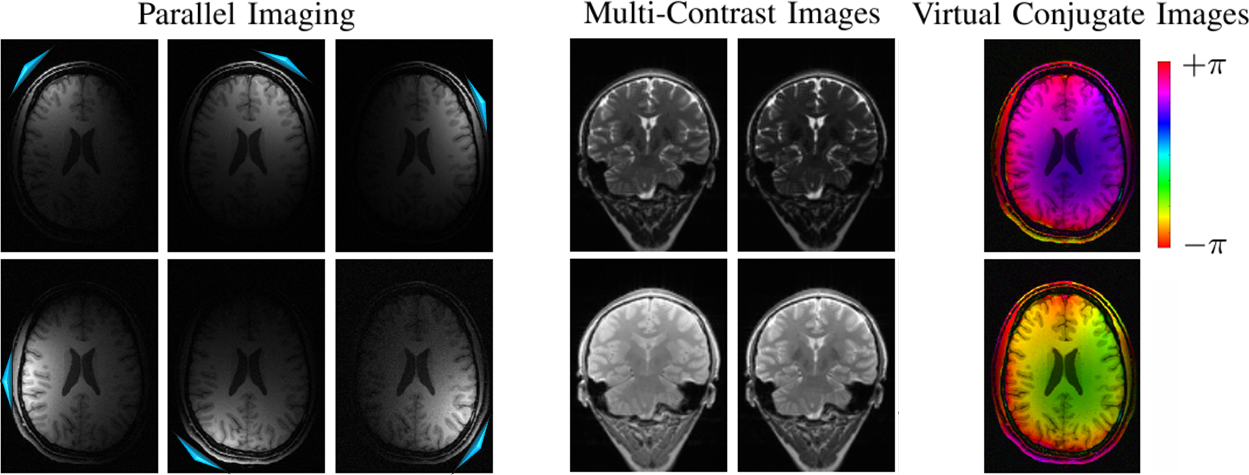}
\caption{Illustration of three typical application scenarios in which multiple MRI images are acquired of the same anatomy.  In the parallel imaging case, we show six images that were obtained simultaneously from an array of receiver coils, where the approximate positions of the coils are indicated in blue. In the multi-contrast case, we show four different brain images that were each acquired with different scan parameters. We show magnitude images for both these cases. In the virtual conjugate case, we show one original image as well as a ``virtual image" obtained by applying complex-conjugation and reversal operations to the measured Fourier data. Since both magnitude images are the same for the virtual conjugate case, we have additionally depicted the image phase with color.\label{fig:multi}}
\end{figure}

There are a variety of practical circumstances that lead to the availability of multiple correlated images:

\begin{itemize}

\item {\bf Parallel Imaging with a Receiver Array.} In this scenario, data is acquired simultaneously from an array of receiver coils \cite{ying2010}.  The images from each coil are all slightly different because each coil has a distinct spatial sensitivity pattern that leads to a coil-specific modulation effect on the image.  

\item {\bf Multi-Contrast Images.}  In this scenario, a sequence of images  is acquired where each image has different contrast characteristics.  These contrast characteristics may vary because the scanner operator has modified some of the scan parameters that influence the way that magnetization evolves under the Bloch equations (i.e., the physical ``equations of motion'' for MRI) to probe the  subject from different spin-physics perspectives, or because the scanner operator has simply acquired images of a time-varying object at different time points.  In this setting, the assumption that $\rho_q(x) = c_q(x)\rho(x)$ for a bandlimited modulation function $c_q(x)$ is sometimes known as the ``generalized series model" in the MRI literature \cite{liang1992}.

\item {\bf Virtual Images using Conjugate Phase Relationships.}  Even in the single-image case where we measure a single dataset $
\tilde{\rho}_1[n]$ corresponding to a single-image $\rho_1(x)$, it is possible to generate a second ``virtual dataset" $\tilde{\rho}_2[n]$ by applying reversal and complex conjugation operations to $\tilde{\rho}_1[n]$  \cite{blaimer2009}, i.e., $\tilde{\rho}_2[n] = \tilde{\rho}_1^*[-n]$, where $^*$ denotes complex conjugation.  Using the symmetry properties of the Fourier transform, it is easy to show that the ``virtual image"  $\rho_2(x)$ corresponding to the virtual data must satisfy $\rho_2(x) = \rho_1^*(x)$.  Note that if $\rho_1(x)$ is decomposed into its magnitude $m(x)$ and phase $\phi(x)$ components as $\rho_1(x) = m(x) e^{i \phi(x)}$, then we can obtain the desired relationship $\rho_q(x) = c_q(x)\rho(x)$ for $q=1,2$ by setting $\rho(x) = m(x)$, $c_1(x) = e^{i \phi(x)}$, and $c_2(x) = e^{-i \phi(x)}$ \cite{haldar2013b}. 

\end{itemize}
All of these multi-image scenarios are illustrated  in Fig.~\ref{fig:multi}.  While  these cases have similarities, it should be noted that sampling considerations are often different in each case.  In parallel imaging, the same sampling pattern must be used for all coils, while the sampling pattern can be different for each image in the multicontrast case.  In the virtual conjugate case, the sampling pattern for one dataset must be the reverse of the pattern for the other dataset.

\subsection{Multi-Image Linear Predictability}\label{sec:multi2}

In the multi-image case, the single-image approximate linear predictability relationship from Eq.~\eqref{eq:interp} can be generalized so that the data sample from one image can be expressed as a  linear combination of   samples from multiple images \cite{sodickson1997,mckenzie2001,griswold2002,lustig2010}:
\begin{equation}
\tilde{\rho}_m[n] \approx \sum_{\substack{q=1\\q\neq  m}}^Q\sum_{k=-L}^{P} \alpha_{qk} \tilde{\rho}_q [n-k]  + \sum_{\substack{k=-L\\k\neq 0}}^{P} \alpha_{mk} \tilde{\rho}_m[n-k]  \text{ for } \forall n\in \mathbb{Z}.\label{eq:extrapm}
\end{equation}
And similar to the derivation of Eq.~\eqref{eq:annihil}, if a set of multiple signals is approximately linearly predictable in the form of Eq.~\eqref{eq:extrapm}, then there must exist coefficients $\{\{\tilde{h}_q[n]\}_{n=-L}^P\}_{q=1}^Q$ with $\tilde{h}_m[0] = -1$ such that
\begin{equation}
0 \approx \sum_{q=1}^Q\sum_{k=-L}^{P} \tilde{h}_q[k] \tilde{\rho}_q [n-k] \text{ for } \forall n \in \mathbb{Z}.\label{eq:annihil2}
\end{equation}

Under what situations does Eq.~\eqref{eq:annihil2} hold?  
The following theorem is a simple generalization of Thm.~1, and  combines ideas from Refs.~\cite{cheung1990,haldar2013b,setsompop2012,haldar2015}:
\begin{theorem}
Assume that the multi-image datasets $\{\tilde{\rho}_q[n]\}_{q=1}^Q$ are defined in terms of $\rho(x)$ and $\{c_q(x)\}_{q=1}^Q$ as described previously, and let $\varepsilon$ be an arbitrary positive  scalar. We have that $$\sum_{n \in \mathbb{Z}} \left|\sum_{q=1}^Q \sum_{k=-L}^{P} \tilde{h}_q[k] \tilde{\rho}_q[n-k]\right|^2 \leq \varepsilon $$ for some set of  filter functions $\{\tilde{h}_q[n]\}_{q=1}^Q$ if and only if  $$ \frac{1}{B}\int_{-B/2}^{B/2} \left|\sum_{q=1}^Q c_q(x) \rho(x) h_q(x)\right|^2 dx \leq \varepsilon, \mathrm{\, where \,}
h_q(x) = \frac{1}{B} \sum_{n=-L}^{P} \tilde{h}_q[n] e^{i 2\pi n x/B}.$$
\end{theorem}
Theorem 2 gives general necessary and sufficient conditions for the approximate annihilability of multichannel data, and has no dependence on the choice of $m$ needed for the linear prediction relationship in Eq.~\eqref{eq:extrapm}.

Recall that in the single-image case, the function $h(x)$ was required to be both sufficiently-smooth and sufficiently large.  In the multi-image case, the functions $h_q(x)$ are still required to be sufficiently-smooth for $q=1,\ldots,Q$.  However, in the multi-image case, approximate linear predictability only requires that $h_m(x)$ is sufficiently large, while the remaining $h_q(x)$ functions with $q\neq m$ are allowed to be arbitrarily small or even zero.

To provide some intuition about this theorem, note that the datasets $\tilde{\rho}_q[n]$ can each be viewed as individual single-image datasets, and therefore each contains intra-image linear prediction and annihilation relationships for the same reasons enumerated in Sec.~\ref{sec:sing}, without the need to involve the other datasets \cite{haldar2015}. In particular, for each $m=1,\ldots,Q$, we can obtain an annihilation relationship that satisfies Thm.~2 by choosing nontrivial coefficients $\{\tilde{h}_m[n]\}_{n=-L}^P$ such that $h_m(x)\rho_m(x) \approx 0$ across the whole FOV, while choosing the remaining $\tilde{h}_q[n]=0$  such that $h_q(x) = 0$ for $q\neq m$.  However, in the multi-image case, it is possible to demonstrate that additional inter-image linear prediction and annihilation relationships exist, and there are a variety of ways to derive theoretical sufficient conditions for these relationships to exist. For example, a sufficient (but not necessary) condition for inter-image relationships can be derived in the case where the $c_q(x)$ are smooth enough that they can be described via bandlimited Fourier series, i.e., $c_q(x) = \frac{1}{B} \sum_{n=-L}^P \tilde{c}_q[n] e^{i2\pi x n /B}$ for $x\in[-B/2,B/2]$ and $q=1,\ldots,Q$.  These relationships can  be demonstrated using theory from the blind multi-channel deconvolution literature, as described elsewhere in this special issue \cite{jacob2019}.  

In this article, we will describe an alternative  perspective that does not require strict bandlimited-modeling of the $c_q(x)$ functions, and is instead motivated by the theory of the SiMultaneous Acquisition of Spatial Harmonics (SMASH) method \cite{sodickson1997,mckenzie2001}.  By definition, the term  appearing on the right side of Eq.~\eqref{eq:annihil2}  can be rewritten as 
\begin{equation}
\sum_{q=1}^Q \sum_{k=-L}^{P} \tilde{h}_q[k] \tilde{\rho}_q[n-k] =\frac{1}{B}\int_{-B/2}^{B/2} \rho(x) e^{-i2\pi n x/B} \left[\sum_{q=1}^Q \sum_{k=-L}^{P} \tilde{h}_q[k]  c_q(x) e^{i2\pi k x/B}\right] dx.
\end{equation}
As a result, if the image $\rho(x)$ has bounded magnitude, we can obtain an inter-image relationship in the form of Eq.~\eqref{eq:annihil2} whenever it is possible to find nontrivial coefficients  $\tilde{h}_q[k]$ for $k=-L,\ldots,P$ and $q=1,\ldots,Q$ such that
\begin{equation}
\sum_{q=1}^Q \sum_{k=-L}^{P} \tilde{h}_q[k]  c_q(x) e^{i2\pi k x/B} \approx 0 \text{ for } \forall x\in [-B/2,B/2].
\end{equation}
For an arbitrarily chosen value of $m\in[1,\ldots,Q]$ and assuming $\tilde{h}_m[0] \neq 0$, this is equivalent to
\begin{equation}
c_m(x) \approx -\frac{1}{\tilde{h}_m[0]}\left(\sum_{\substack{q=1\\q\neq m}}^Q \sum_{k=-L}^P \tilde{h}_q[k]c_q(x) e^{i2\pi k x/B} + \sum_{\substack{k =-L\\k\neq 0}}^P \tilde{h}_m[k] c_m(x)e^{i2\pi k x/B} \right) \text{ for } \forall x\in [-B/2,B/2].
\end{equation}
Thus, the conditions of Thm.~2 can be satisfied whenever one modulation function $c_m(x)$ can be  well approximated by a linear combination of harmonically-modulated versions of $\{c_q(x)\}_{q=1}^Q$ \cite{mckenzie2001}.  Note however that this  is only sufficient (not necessary) for linear predictability, in contrast to the necessary and sufficient conditions of Thm.~2. 

\subsection{Simultaneous multi-slice imaging}
While the previous subsections described linear predictability in classical MRI settings with traditional data acquisition, a more advanced data acquisition technique known as simultaneous multi-slice imaging has recently become increasingly popular, with a tremendous practical impact on modern MRI experiments.  While the low-level details of this approach are beyond the scope of this article (see Ref.~\cite{barth2016} for an in-depth treatment), we will present a high-level perspective of this approach that is sufficient  to make connections to linear prediction apparent.

\begin{figure}[htp]
\floatbox[{\capbeside\thisfloatsetup{capbesideposition={right,center},capbesidewidth=3.5in}}]{figure}[\FBwidth]{
\caption{An illustration of simultaneous multi-slice imaging, in which Fourier data $\tilde{s}[n]$ is collected corresponding to an image $s(x)$ that is  a superposition of multiple images of interest (in this case, the two images $\rho_1(x)$ and $\rho_2(x)$).  The goal is to recover the Fourier data for the individual images from a potentially-undersampled version of the superposed data $\tilde{s}[n]$. }
\label{fig:sms}}{
\begin{minipage}{3in}
\centering
\begin{minipage}{0.8in}
\centering 
\footnotesize $s(x)$ \\[0.25em]
\includegraphics[width=.8in]{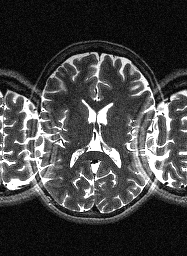}
\end{minipage}
\begin{minipage}{0.8in}
\centering 
\footnotesize $\rho_1(x)$  \\[0.25em]
\includegraphics[width=.8in]{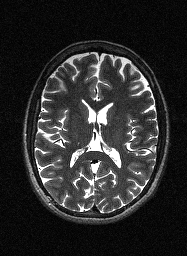}
\end{minipage}
\begin{minipage}{0.8in}
\centering 
\footnotesize $\rho_2(x)$ \\[0.25em]
\includegraphics[width=.8in]{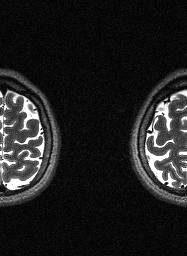}
\end{minipage}
\end{minipage}}
\end{figure}

In simultaneous multi-slice imaging, the MRI data acquisition physics are manipulated so that we observe Fourier data simultaneously from the superposition of multiple images, as illustrated in Fig.~\ref{fig:sms}.  For example, assuming   we observe $R$ such simultaneous images $\{\rho_r(x)\}_{r=1}^R$, the data acquisition can be modeled as
\begin{equation}
\tilde{s}[n] = \int_{-B/2}^{B/2} s(x) e^{-i2\pi n x/B} dx \mathrm{\,   where  \, } s(x) = \sum_{r=1}^R \rho_r(x).\label{eq:super}
\end{equation}
The goal  is then to recover the collection of individual images $\{\rho_r(x)\}_{r=1}^R$ (or equivalently, the samples of their Fourier transforms $\{\tilde{\rho}_r[n]\}_{r=1}^R$) from a potentially-undersampled version of $\tilde{s}[n]$.\footnote{Note that we are describing a simplified version of this approach to avoid complicated notation -- in most practical applications, simultaneous multi-slice data will leverage data acquired from an array of receiver coils \cite{barth2016}.}  The simultaneous multi-slice approach has been able to have such a major impact on MRI because it allows the acquisition of multiple images in the same amount of time that would normally be required to acquire a single image.  

Importantly, many simultaneous multi-slice reconstruction methods assume that  the Fourier data of the individual images $\{\tilde{\rho}_r[n]\}_{r=1}^R$ can be recovered as a linear shift-invariant combination of the samples $\tilde{s}[n]$ \cite{setsompop2012,barth2016}.  For instance, with an arbitrarily selected value of $m \in [1,\ldots,R]$, a typical assumption might be that coefficients $\{\alpha_{mk}\}_{k=-L}^P$ exist such that
\begin{equation}
\tilde{\rho}_m[n] \approx \sum_{k = -L}^P \alpha_{mk} \tilde{s}[n-k] \text{ for } \forall n\in \mathbb{Z}.\label{eq:sms1}
\end{equation}
In practice, it can also be helpful to impose additional constraints to avoid interslice leakage \cite{cauley2014}.  For example,  in addition to imposing the prediction relationship from Eq.~\eqref{eq:sms1}, we can help to avoid contributions from other slices leaking into the reconstruction of the $m$th slice by additionally imposing that
\begin{equation}
0 \approx \sum_{k=-L}^P \alpha_{mk} \tilde{\rho}_r[n-k] \text{ for } \forall n\in \mathbb{Z}  \text{ and for } \forall r \neq m,\label{eq:sms4}
\end{equation}
which can clearly be interpreted as annihilation relationships.

How can we justify and understand Eq.~\eqref{eq:sms1}?  Using the fact that $\tilde{s}[n] = \sum_{r=1}^R \tilde{\rho}[n]$, it is straightforward to rewrite Eq.~\eqref{eq:sms1} in the form of an annihilation relationship
\begin{equation}
0\approx -\tilde{\rho}_m[n]+\sum_{r=1}^R \sum_{k = -L}^P \tilde{h}[k]  \tilde{\rho}_r[n-k] \text{ for } \forall n\in \mathbb{Z},\label{eq:sms3}
\end{equation}
where $\tilde{h}[k] = \alpha_{mk}$ for all $k=-L,\ldots,P$.  This annihilation relationship is similar to the previous ones, and is associated with the following theorem (representing a combination of arguments from Refs.~\cite{cheung1990,haldar2013b,setsompop2012,haldar2015}):
\begin{theorem}
Assume that $\tilde{s}[n]$, $\{\tilde{\rho}_r[n]\}_{r=1}^R$, and $\{\rho_r(x)\}_{r=1}^R$ are defined as described previously, and let $\varepsilon$ be an arbitrary positive  scalar. We have that $$\sum_{n \in \mathbb{Z}} \left|-\tilde{\rho}_m[n]+\sum_{r=1}^R \sum_{k=-L}^{P} \tilde{h}[k]  \tilde{\rho}_r[n-k]\right|^2 \leq \varepsilon $$ for some  filter  $\tilde{h}[n]$ if and only if  $$ \frac{1}{B}\int_{-B/2}^{B/2} \left| -\rho_m(x)+ h(x) \sum_{\substack{r=1}}^R \rho_r(x)\right|^2 dx \leq \varepsilon, \mathrm{\, where \,}
h(x) = \frac{1}{B} \sum_{n=-L}^{P} \tilde{h}[n] e^{i 2\pi n x/B}.$$
\end{theorem}
Achieving Eq.~\eqref{eq:sms1} thus requires that it is possible to find a smooth function $h(x)$ such that $h(x)\sum_{\substack{r=1}}^R \rho_r(x) \approx \rho_m(x)$ within the FOV.  If this result is combined with the constraints from Eq.~\eqref{eq:sms4} (which require that the same smooth function also satisfies $h(x) \rho_r(x) \approx 0$ for each $r \neq m$), then we must simultaneously have that both $h(x) \approx 1$ within the support of $\rho_m(x)$ and $h(x) \approx 0$ within the supports of each $\rho_r(x)$ for all $r \neq m$.  Clearly, this requires that the approximate support of $\rho_m(x)$ is disjoint from the approximate supports of  $\rho_r(x)$ for all $r \neq m$. While this support condition is quite restrictive (e.g., it does not hold for the example shown in Fig.~\ref{fig:sms}), it should be noted that careful manipulation of the imaging physics allows substantial control over the superposition relationship in Eq.~\eqref{eq:super}, and that some acquisition schemes will satisfy this condition better than others \cite{setsompop2012,barth2016}.  In addition, simultaneous multi-slice imaging is usually used in settings where data is acquired simultaneously using an array of receiver coils, which leads to a large multiplicity of inter-image annihilation relationships  for the same reasons described in Section~\ref{sec:multi2}.  These inter-image annihilation relationships enable linear prediction-based reconstruction across a much broader range of imaging contexts.  The derivations for the multi-coil simultaneous multi-slice case are relatively straightforward extensions of our previous descriptions, and we omit them.

\section{Application Examples}

\begin{figure}[htp]
\floatbox[{\capbeside\thisfloatsetup{capbesideposition={right,center},capbesidewidth=3.5in}}]{figure}[\FBwidth]{
\caption{A 3D rendering of a  reconstruction result from Ref.~\cite{kim2018c}.  Using linear predictability and structured low-rank matrix modeling to impose support, phase, and multi-image (parallel imaging) constraints, this detailed high-quality in vivo human head image was able to be reconstructed from roughly 40 seconds worth of calibrationless data,  which is a substantial 16-fold improvement over the $>$10 minute acquisition required to sample data fully at the Nyquist rate. \label{fig:brain}}}{
\includegraphics[width=1.5in]{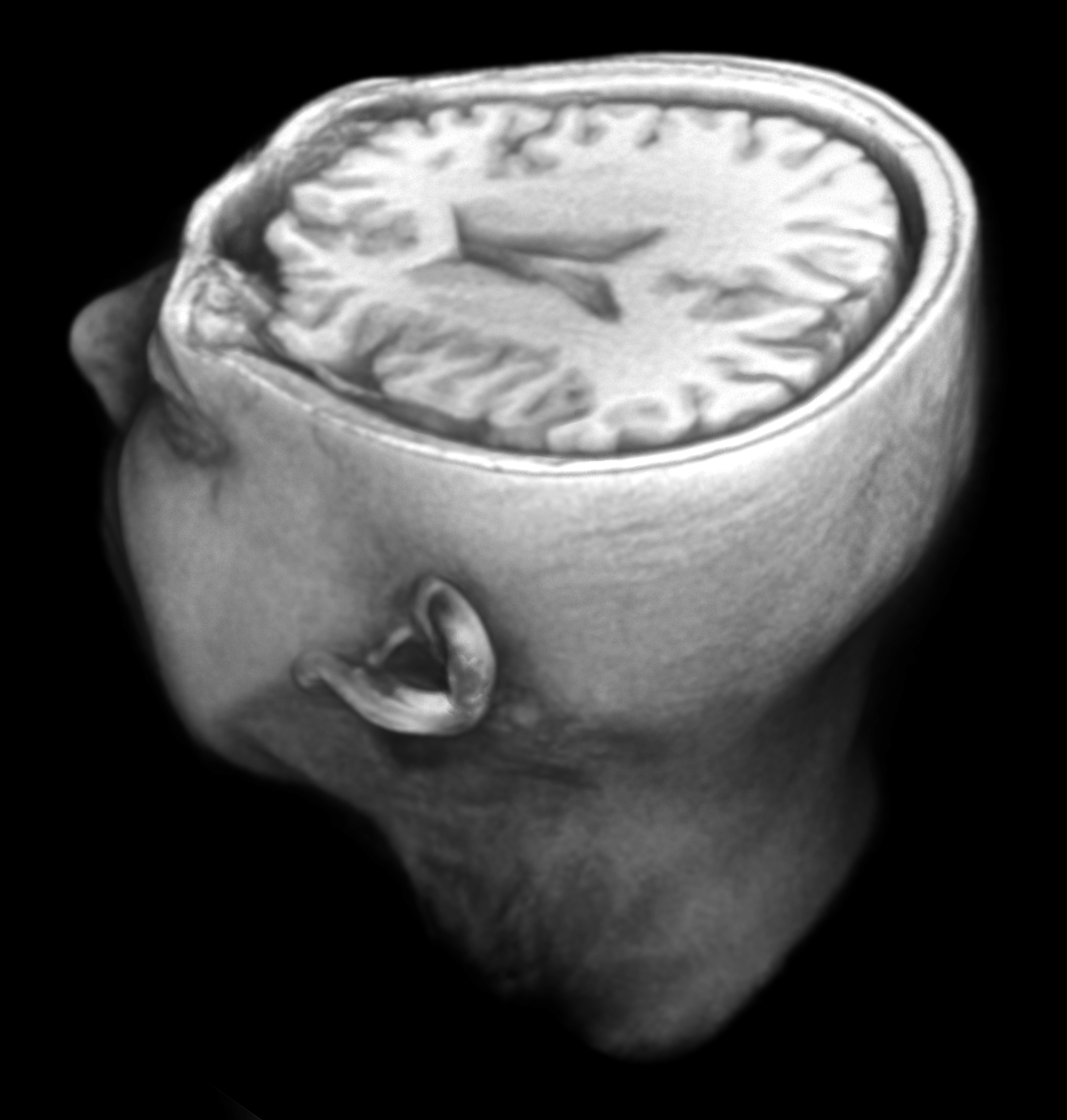}}
\end{figure}

Linear prediction-based reconstruction methods have had a major impact across a broad range of MRI application domains, and in this section, we present a few example illustrations.  Our first example relates to high-resolution $T_1$-weighted 3D brain imaging, which provides a detailed view of the macroscopic anatomy of the brain, with a clear delineation between tissues like white  and gray matter.  This kind of image has some clinical relevance, although is perhaps most widely acquired for neuroscience studies. In neuroscience, this type of image is routinely used as the basis for structural alignment and anatomical correspondence between different images, and  is also commonly used to measure morphological features (i.e., the volumes, thicknesses, etc. of different brain structures)  that can be neuroscientifically relevant because they often vary with developmental stage, intelligence, health history, etc.  While this kind of image is very useful, it can also be time-consuming to acquire if data is sampled at the conventional Nyquist rate.  For example, imaging the entire brain with (1 mm)$^3$ isotropic resolution can frequently require more than 10 minutes using conventional Nyquist sampling.  Figure~\ref{fig:brain} depicts a reconstruction from Ref.~\cite{kim2018c} in which linear predictability (specifically, structured low-rank matrix completion within the LORAKS framework \cite{haldar2013b,haldar2015,haldar2015b}) enables an image to be reconstructed from 16-fold undersampled data, corresponding to an acquisition time of only $\sim$40 seconds.  This level of acceleration is very significant for a variety of reasons, including the fact that shorter scans are associated with fewer motion artifacts.  For example, it is relatively easy for most human subjects to stay still inside the scanner for a 40 second experiment, but many subjects struggle to stay still over 10-minute spans!  In addition, this dramatic improvement in data acquisition time enables the scanner operator to either image a larger number of subjects or image each subject with a higher level of detail within a fixed acquisition duration.

\begin{figure}[htp]
\includegraphics[width=5.6in]{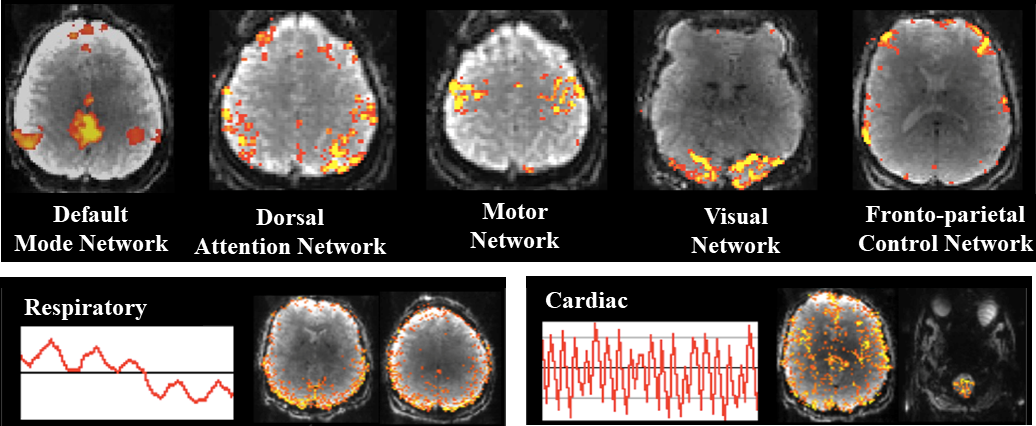}
\caption{Using simultaneous multi-slice imaging with $R=12$ simultaneous slices and a calibrated linear prediction-based reconstruction   that embeds simultaneous multi-slice, support, and multi-image (parallel imaging) constraints, it is possible to acquire whole-brain images with (2.5 mm)$^3$ isotropic resolution in just 350 ms \cite{setsompop2012}.  This enables the high-quality estimation of resting-state functional connectivity network information from a relatively short experiment, since a short 5-minute scan can provide roughly 900 time points of data.  In addition, this rapid imaging also enables a temporal sampling rate that satisfies the Nyquist rate for fast physiological nuisance signals (e.g., respiratory and cardiac signals).  While these nuisance signals would normally alias and confound the interpretation of brain activity for conventional slower scanning methods, the rapid imaging a protocol now allows these signals to be cleanly removed.  }\label{fig:fmri}
\end{figure}

Our second example relates to functional MRI (fMRI), in which a time-resolved sequence of MRI images is acquired to provide insight into the dynamic neural activity that occurs in the brain while subjects are resting, responding to stimuli, or performing tasks.   The use of linear prediction-based reconstruction methods has recently enabled a revolution in the achievable spatiotemporal resolution of fMRI experiments.  For example, Fig.~\ref{fig:fmri} shows a case in which simultaneous multi-slice data acquisition with linear prediction-based image reconstruction enables 12-fold acceleration of the data acquisition time for each 3D image \cite{setsompop2012}, allowing an entire brain volume to be acquired in just 350 ms.  This temporal resolution makes it much easier to identify brain activity and functional connectivity network information, providing potentially profound new insights into  human brain function and malfunction. 

\begin{figure}[htp]
\includegraphics[width=4.7in]{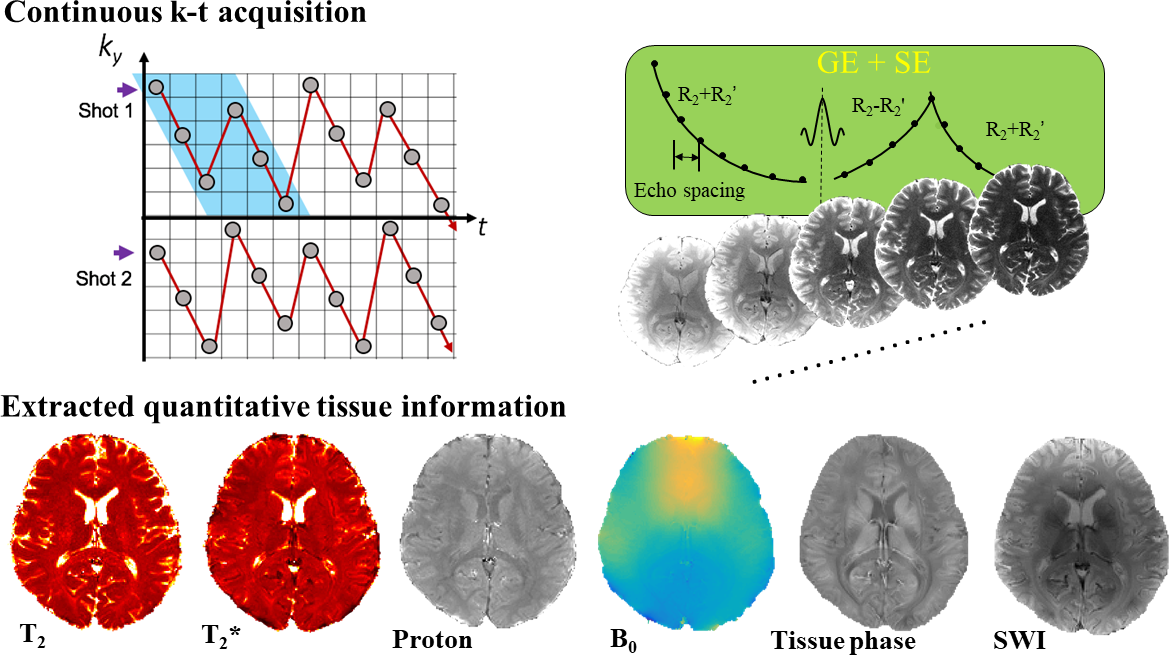}
\caption{An illustration of Echo Planar Time-resolved imaging (EPTI) \cite{wang2019}, in which a series of multi-contrast images are acquired using a novel multi-shot continuous spatiotemporal $(k,t)$ acquisition strategy, and missing data is linearly interpolated from the series of multi-contrast datasets.  Using calibration data, the reconstruction incorporates simultaneous multi-slice and multi-image (parallel imaging) constraints, as well as implicitly incorporating spatial-spectral support constraints (i.e., following the same argument from Thm. 1, linear predictability exists in the $(k,t)$ domain for spatiotemporal images that obey a support constraint in the reciprocal $(x,f)$ space \cite{haldar2013b}, where $f$ is the frequency variable corresponding to time $t$).  This approach allows the reconstruction of hundreds of high-fidelity images with different contrast weightings, which can be generated at a very high temporal sampling rate of $\sim$1 ms. Once these images are reconstructed, they can then be used to extract quantitative tissue parameter maps that provide detailed information related to microstructural tissue features.}
\label{fig:epti}
\end{figure}

Our third and final example relates to acquiring a sequence of multi-contrast MRI images that can be used to extract a variety of important quantitative tissue parameters.  While multi-contrast imaging is usually very time consuming, a novel data acquisition approach enabled by linear prediction-based reconstruction (with simultaneous multi-slice imaging, parallel imaging with an array of receiver coils, and spatiotemporal image reconstruction) enables hundreds of whole-brain images to be acquired in less than 30 seconds \cite{wang2019}, which is an acceleration factor of 50-80$\times$ relative to  Nyquist sampling. This case is illustrated in Fig.~\ref{fig:epti}.

\section{Summary and Outlook}

As we have described in this article, Fourier MRI data can be shown to possess autoregressive structure for a variety of distinct reasons, and it is relatively straightforward to determine when such structure will or will not exist.  When this structure is present, it implies that computational MRI reconstruction methods based on shift-invariant linear prediction or annihilation can be leveraged to enable imputation of the missing information from highly-undersampled data acquisitions.  Various kinds of linear prediction-based methods have been studied in MRI over the past several decades, and some of the most powerful and influential modern computational MRI reconstruction methods are deeply tied to linear prediction principles.  However, although these ideas have already been studied for a long time, there are still many important open questions that still need to be addressed, and recent years have witnessed the emergence of a variety of important new theoretical insights and fresh new computational imaging formulations and algorithms.  At the same time, there has been an increasing transference of these  ideas from academic research into practical clinical and scientific MRI applications, where the power of linear predictability is enabling new experimental paradigms that are pushing the frontiers of what we can observe with MRI.  It should also be noted that while our description was specific to the context of MRI, many of the constraints we have  used (i.e., limited support, smooth phase, correlation between multiple images, transform-domain sparsity) are quite general and occur commonly in many other inverse problems. As a result, we believe that the association between these constraints and Fourier-domain autoregression and linear predictability also has a strong potential to enable reconstruction improvements in a wide range of application settings beyond MRI.

\section*{Acknowledgments}
This work was supported in part by research grants NSF CCF-1350563, NIH R21-EB022951, NIH R01-MH116173, NIH R01-NS074980, NIH R01-NS089212, NIH R33-CA225400, and NIH R01-EB020613.  The authors are grateful to Antonio Ortega, Krishna Nayak, Richard Leahy, and Rodrigo Lobos for valuable comments.

\singlespacing
\bibliographystyle{IEEEtran}
\bibliography{./bibliography}

\end{document}